\documentclass[prd,letterpaper,onecolumn,amsmath,amsfonts,amssymb,nofootinbib]{revtex4}
\usepackage[dvips]{graphicx}
\usepackage{psfrag}
\usepackage{bm}
\usepackage{dcolumn}
\begin{document}
\title{Cosmological Evolution of Homogeneous Universal Extra Dimensions}
\author{Torsten Bringmann}
\email{troms@physto.se}
\author{Martin Eriksson} 
\email{mate@physto.se} 
\author{Michael Gustafsson}
\email{michael@physto.se}
\affiliation{Department of Physics, AlbaNova, Stockholm University, SE
  - 106 91 Stockholm, Sweden} 
\date{September 30, 2003}
\begin{abstract}
The lightest Kaluza-Klein particle appearing in models with universal
extra dimensions has recently been proposed as a viable dark matter
candidate when the extra dimensions are compactified on a scale of the
order of 1~TeV. Underlying assumptions of this proposal are that the
size of the extra dimensions stays constant and that the evolution of
the universe is given by standard cosmology. Here we investigate, both
analytically and numerically, whether this is possible without
introducing an explicit stabilization mechanism. By analyzing
Einstein's field equations for a
\mbox{($3\!+\!n\!+\!1$)}\,-\,dimensional homogeneous, but in general
anisotropic universe, we find that approximately static extra
dimensions arise naturally during radiation domination. For matter
domination, however, there are no solutions to the field equations that
allow static extra dimensions or the usual behavior of the scale
factor for ordinary three-dimensional space. We conclude that an
explicit mechanism is needed in order to stabilize the extra
dimensions and reproduce standard cosmology as we know it.
\end{abstract}
\maketitle

\newcommand{\dims}{\mbox{($3\!+\!1$)\,}}
\newcommand{\ndims}{\mbox{($3\!+\!n\!+\!1$)\,}}
\newcommand{\be}{\begin{equation}}
\newcommand{\ee}{\end{equation}}
\newcommand{\da}{\dot{a}}
\newcommand{\db}{\dot{b}}
\newcommand{\dn}{\dot{n}}
\newcommand{\dda}{\ddot{a}}
\newcommand{\ddb}{\ddot{b}}
\newcommand{\ddn}{\ddot{n}}
\newcommand{\pa}{a^{\prime}}
\newcommand{\pb}{b^{\prime}}
\newcommand{\pn}{n^{\prime}}
\newcommand{\ppa}{a^{\prime \prime}}
\newcommand{\ppb}{b^{\prime \prime}}
\newcommand{\ppn}{n^{\prime \prime}}
\newcommand{\fda}{\frac{\da}{a}}
\newcommand{\fdb}{\frac{\db}{b}}
\newcommand{\fdn}{\frac{\dn}{n}}
\newcommand{\fdda}{\frac{\dda}{a}}
\newcommand{\fddb}{\frac{\ddb}{b}}
\newcommand{\fddn}{\frac{\ddn}{n}}
\newcommand{\fpa}{\frac{\pa}{a}}
\newcommand{\fpb}{\frac{\pb}{b}}
\newcommand{\fpn}{\frac{\pn}{n}}
\newcommand{\fppa}{\frac{\ppa}{a}}
\newcommand{\fppb}{\frac{\ppb}{b}}
\newcommand{\fppn}{\frac{\ppn}{n}}
\newcommand{\fkaa}{\frac{k_a}{a^2}}
\newcommand{\fkbb}{\frac{k_b}{b^2}}
\newcommand{\fub}{\frac{db}{du}}
\newcommand{\fuub}{\frac{d^2 b}{du^2}}
\newcommand{\deltaa}{\delta_a}
\newcommand{\deltab}{\delta_b}
\newcommand{\deltarho}{\delta_\rho}
\newcommand{\ddeltaa}{\dot{\delta}_a}
\newcommand{\ddeltab}{\dot{\delta}_b}
\newcommand{\dddeltaa}{\ddot{\delta}_a}
\newcommand{\dddeltab}{\ddot{\delta}_b}

\section{Introduction}
Everyday experience seems to suggest that our world consists of four
space-time dimensions. However, at the beginning of the
$20^\text{th}$ century Nordstr\"om, Kaluza, and Klein (KK) already realized
that this may in fact not be the case \cite{nor,kal,kle}. In the last few
years there has again been a great deal of interest in models with extra
dimensions, most notably due to the influence of string (or M) theory
which in its usual formulation requires more than four dimensions
(see e.g.~\cite{pol}). In particular, this has led to a number of brane-world
scenarios where all, or only some, of the gauge bosons are allowed to
propagate in the extra dimensions while matter fields are restricted to
\dims -\,dimensional branes. For a review of different models with
extra dimensions, see, for example, \cite{hew} and references therein.

In this context, a specific model of so-called universal extra
dimensions (UED) has recently been proposed by Appelquist \textit{et
al.} \cite{appa}, in which \textit{all} standard model fields are
allowed to propagate in the extra dimensions. As usual, quantization
of the extra-dimensional momentum leads to a tower of KK states that
appear as new massive particles in the effective four-dimensional
theory. The existing constraints on electroweak observables translate
into bounds on the compactification scale $R$, which is related to the
mass of the lowest excitations by $M \sim 1/R$. For one or two
UED these bounds are of the order of a few hundred GeV
and thus within reach of the Large Hadron Collider (LHC) or the
Fermilab Tevatron run II \cite{appa,riz,cheng,che}.

The UED model is not only of great interest from the point of view of
particle physics \cite{arka,dob,appb,moha}, but might also
provide a solution to one of the most outstanding puzzles in modern
cosmology -- the nature of dark matter \cite{sla,cheng2,sera}. In the UED
scenario the lightest KK particle (LKP) is stable because of KK parity
conservation and could therefore still be present today as a thermal
relic. Furthermore, if it is also neutral and nonbaryonic it has all
the properties of a weakly interacting massive particle (WIMP), one of
the most promising dark matter candidates (see \cite{jung} for a nice
introduction to WIMP dark matter). According to \cite{sera}, both the
KK photon ($B^{(1)}$) and the KK neutrino could account for dark
matter with $\Omega_M \sim 0.3$, as suggested by the current
cosmological concordance model \cite{map}, if one assumes a
compactification scale of about $R \sim 1~\textrm{TeV}^{-1}$
size. Indirect and direct detection properties of such KK dark matter
candidates are promising for the next generation detectors
\cite{serb,hoo,cheb,ber}. In fact, the KK neutrino seems to be ruled
out already by present data \cite{serb}.

However, when considering the freeze-out process and the further
cosmological evolution of thermally produced LKPs, the size of the
extra dimensions has so far been assumed to stay constant -- although no
explicit stabilization mechanism has been given. As already noted in
\cite{sai,sera}, the resulting relic density today depends crucially
on this assumption, and it is therefore important to investigate
whether it can be justified within the UED framework. Since the UED
may be relatively large (in fact, they must be if the LKPs are not to
overclose the universe), their evolution should be governed by
Einstein's field equations. The aim of this work is therefore to
carefully study the dynamics of the universe as described by an
appropriate extension of the usual Friedmann equations that results
from the field equations in higher dimensions. Specifically, we focus
on solutions with constant or only slowly varying extra dimensions in
the absence of any explicit stabilization mechanism. However, our
analysis is more general in that we investigate whether there is
\emph{any} time evolution of the extra dimensions that corresponds to
standard cosmology in four dimensions; in this context, we also
include a numerical study of the transition regime between the
radiation- and matter-dominated eras. For earlier work on the
stability properties of higher-dimensional cosmologies, often referred
to as Kaluza-Klein or multidimensional cosmology, see, for example,
\cite{fre,appkk,daemi,mae,mae2,gun,gun2,car,cha} and references
therein.

This paper is organized as follows: In Sec.~\ref{ued} we introduce
the cosmological solutions to Einstein's field equations for a
\ndims-\,dimensional homogeneous, but in general anisotropic,
universe. Here, we also comment on the interpretation of pressure in
higher dimensions and derive a general relation between pressure and
energy density in UED cosmology. Necessary conditions that every
solution with static extra dimensions in such a model must satisfy are
then derived in Sec.~\ref{static}. In the next two sections we
study the existence of solutions with (nearly) constant extra
dimensions during radiation and matter domination, respectively. A
possible transition between these two regimes is then outlined in
Sec.~\ref{trans}. Finally, Sec.~\ref{conc} discusses the
implications for the UED scenario and concludes.

\section{Setup}
\label{ued}
\subsection{Basic equations}
We introduce $n$ universal extra dimensions  and adopt
coordinates $X^A$, $A = 0, 1, \ldots, 3 + n$, with
\begin{equation}
  x^\mu \equiv X^\mu \quad (\mu = 0, 1, 2, 3) \,
\end{equation}
and
\begin{equation}
  x^i \equiv X^i \quad (i = 1, 2, 3) \,
\end{equation}
being the coordinates for ordinary four-dimensional spacetime and
three-dimensional (3D) space respectively, and
\begin{equation}
  y^p \equiv X^{3+p} \quad (p = 1, \ldots, n) \,
\end{equation}
the coordinates for the UED. In the absence of a cosmological constant,
Einstein's field equations are given by
\begin{equation}
  G^A_{\phantom{A}B} \equiv R^A_{\phantom{A}B} - \frac{1}{2} R\,
  \delta^A_{\phantom{A}B} = \kappa^2 T^A_{\phantom{A}B} \,. \label{fe} 
\end{equation}
Here, $\kappa^2$ is defined as
\begin{equation}
  \kappa^2 = \frac{8 \pi}{M^{2 + n}} \,, 
\end{equation}
where $M$ is the higher-dimensional Planck mass. In the case of
compactified extra dimensions with volume $V_{(n)}$ it is related to
the usual Planck mass by \cite{arkc}
\begin{equation}
  \label{mpl}
  M_{Pl}^2=V^{\phantom{+}}_{(n)}M^{2+n} \,.
\end{equation}

In the UED scenario, there is no localization mechanism that confines
particles to a brane, so we assume that the energy density is
distributed homogeneously throughout all dimensions. Thus, we are
looking for homogeneous solutions to the field equations that are
isotropic in ordinary three-dimensional space and also -- but
separately -- in the space of extra dimensions. This can be described
by the standard Friedmann-Robertson-Walker (FRW) metric if we allow
for different scale factors in 3D and the UED:
\begin{equation}
  ds^2 = -dt^2 + a^2(t) \gamma_{ij} dx^i dx^j + b^2(t) \tilde{\gamma}_{pq}
  dy^p dy^q \,, \label{metric} 
\end{equation}
where $\gamma_{ij}$ and $\tilde{\gamma}_{pq}$ are maximally symmetric
metrics in three and $n$ dimensions, respectively. Spatial curvature
is thus parametrized in the usual way by $k_a = -1,0,1$ in ordinary
space and $k_b = -1,0,1$ in the UED. Of course, one could imagine a
model that is not described by the metric (\ref{metric}). For
instance, there is no theoretical or observational argument against
having separate scale factors for each extra dimension. We choose this
metric because it is the simplest realistic alternative for studying
dynamical extra dimensions.

With our choice of metric, the energy-momentum tensor must take the
following form:
\begin{equation}
\label{emtensor} 
  T^A_{\phantom{A}B} = \left( 
  \begin{array}{ccc}
    -\rho & 0 & 0 \\
    0 & \gamma^i_{\phantom{i} j} p_a & 0 \\
    0 & 0 & \tilde{\gamma}^p_{\phantom{p} q} p_b
  \end{array}
  \right) \,,
\end{equation}
which describes a homogeneous but in general anisotropic perfect fluid
in its rest frame. The pressure in ordinary space (UED) is related
to the energy density by an equation of state $p_a = w_a \rho$ ($p_b =
w_b \rho$).

The nonzero components of the field equations~(\ref{fe}) are then given by
\begin{subequations}
\label{allfe}
\begin{eqnarray}
  3\left( \fda \right)^2 + 3\fkaa + 3 n \fda \fdb + \frac{n(n -
    1)}{2}\left[\left( \fdb \right)^2 + \fkbb \right] &=&
  \kappa^2\rho \,, \label{fe00} \\
  2\fdda + \left( \fda \right)^2 + \fkaa + n \fddb + 2n \fda \fdb +
  \frac{n(n - 1)}{2} \left[ \left( \fdb\right)^2 + \fkbb \right] &=&
  -\kappa^2 w_a \rho \,, \label{feij}\\
  3\fdda + 3\left( \fda \right)^2 + 3\fkaa + (n-1)\fddb
  +3(n-1)\fda\fdb + \frac{(n - 1)(n -2)}{2}\left[ \left( \fdb \right)^2 + \fkbb
  \right] &=&-\kappa^2 w_b \rho \,, \label{fepq} 
\end{eqnarray}
\end{subequations}
where an overdot denotes differentiation with respect to cosmic time
$t$. From conservation of energy $T^A_{\phantom{A}0;A\phantom{A}} =
0$ we find, furthermore,
\begin{equation}
  \frac{\dot{\rho}}{\rho} = -3(1 + w_a) \fda - n(1 + w_b) \fdb \,.
  \label{econserv}
\end{equation}
For constant equations of state this can be integrated to give
\begin{equation}
  \rho = \rho_i \left( \frac{a}{a_i} \right)^{-3(1 + w_a)} \left(
  \frac{b}{b_i} \right)^{-n(1 + w_b)} \,. \label{rho} 
\end{equation}
We will use a subscript $i$ to indicate arbitrary initial values
throughout.

\subsection{On energy density and pressure}
The energy density and pressure appearing in the above equations are
not the usual three-dimensional quantities but their
higher-dimensional analogues. The pressure in some direction $X^A$ is
conventionally defined as the momentum flux through hypersurfaces of
constant $X^A$. This can be expressed as
\begin{equation}
  p_A = \left< \frac{\mathbf{k}_A^2}{E} \right> \equiv g\int
  \frac{\mathbf{k}_A^2}{E} f(\mathbf{k},\mathbf{x},t) \,\textrm{d}^{3+n}
  \mathbf{k} \,, \label{pressure}
\end{equation}
where $\mathbf{k}_A$ is the momentum in direction $X^A$, $g$ is the
statistical weight and $f(\mathbf{k},\mathbf{x},t)$ gives the phase
space probability distribution. Isotropy in our model means that $p_a
= \left<\mathbf{k}_1^2/E \right> = \left<
\mathbf{k}_2^2/E \right> = \left< \mathbf{k}_3^2/E
\right>$ and $p_b = \left< \mathbf{k}_4^2/E \right> = \ldots =
\left< \mathbf{k}_{n+3}^2/E \right>$. Therefore we find
\begin{equation}
  3p_a + np_b = \rho - \left< \frac{m^2}{E} \right> \, \label{eos_gen}
\end{equation}
where $\rho = \left< E \right>$ and $m$ is the mass of the particles
producing the pressure. In the case of different particle species one
has to sum over all of them in Eq.~(\ref{pressure}), and the mass appearing in
Eq.~(\ref{eos_gen}) can then be interpreted as the effective mass of all
particles. For highly relativistic particles, Eq.~(\ref{eos_gen}) reduces to
\begin{equation}
  3w_a + nw_b = 1 . \label{eos}
\end{equation}
Setting $w_a = w_b$ (corresponding to a completely isotropic \ndims -
dimensional universe) would then result in the equation of state
\begin{equation}
  p = \frac{\rho}{3 + n} \,. \label{n_rad_dom}
\end{equation}
As expected, for $n = 0$ we find the well-known relation for a
relativistic gas in \dims dimensions.

How can we recover standard cosmology with this setup? Let us first
consider the case of highly relativistic particles and static, compact
extra dimensions. Equations (\ref{fe00}) and (\ref{feij}) are then
equivalent to the ordinary Friedmann equations with three-dimensional
energy density $\rho^{(3)}$, pressure $p^{(3)}$, and an effective cosmological
constant $\Lambda_{\textrm{eff}}$ given by
\begin{subequations}
\begin{eqnarray}
  \rho^{(3)} &=& V_{(n)} \rho \,, \\
  p^{(3)} &=& w_a \rho^{(3)} \,, \\
  \Lambda_{\textrm{eff}} &=& -\frac{n(n - 1)}{2} \frac{k_b}{b^2} \,. 
\end{eqnarray}
\end{subequations}
Moreover, from Eq.~(\ref{rho}) we then find the standard cosmological
evolution of $\rho^{(3)} \propto a^{-3(1 + w_a)}$. For vanishing
extra-dimensional curvature, all we need in order to recover the
familiar case of \dims-\,dimensional radiation domination is to set $w_a
= 1/3$. However, this forces us to allow for different pressures in
ordinary space and the UED, since according to Eq.~(\ref{eos}) $w_b$
\textit{must} then be close to zero. That is, the
extra dimensional pressure (and momentum) must be negligible, and thus the KK tower must be largely unpopulated compared to the 3D radiation in order
to reproduce standard cosmology for a radiation-dominated stage with $w_a\approx \frac{1}{3}$.

We also note that if LKPs do form a substantial part of the dark
matter, then from a \ndims-\,dimensional point of view (ignoring a
possible epoch of vacuum energy domination) the universe is
\textit{always} dominated by relativistic particles. This is because
any standard model particle with extra-dimensional momentum is
automatically relativistic since $m \ll 1/R\sim
1~\textrm{TeV}$. During ordinary radiation domination, on the other
hand, the contribution from dark matter is negligible -- but then the
dominant part of the energy density is relativistic anyway. Therefore,
Eq.~(\ref{eos}) is always valid in our model. Of course, we still want
$w_a\approx0$ in order to describe what looks like a 3D matter
dominated universe, so we have to set
\begin{equation}
  w_b\approx\frac{1}{n} \, \label{md_wb}
\end{equation}
in that case.

On the other hand, for a time-dependent scale factor $b$,
Eqs.~(\ref{fe00}) and~(\ref{feij}) can still be cast in the standard
cosmological form by absorbing all terms containing factors of $b$ and
its derivatives into an \textit{effective} three-dimensional energy
density and pressure:\footnote{For an alternative definition of the
effective pressure, see \cite{mu}.}
\begin{subequations}
\label{effective}
\begin{eqnarray}
  \rho_{\textrm{eff}}^{(3)} &=& \frac{M_{Pl}^2}{8 \pi} \left\{
  \kappa^2 \rho - 3 n \fda \fdb - \frac{n(n - 1)}{2}\left[\left( \fdb
  \right)^2 + \fkbb \right] \right\} \,, \\ 
  p_{\textrm{eff}}^{(3)} &\equiv&  w_{\textrm{eff}}
  \rho_{\textrm{eff}}^{(3)} = \frac{M_{Pl}^2}{8 
  \pi} \left\{ \kappa^2 w_a \rho + n \fddb + 2n \fda \fdb +
  \frac{n(n - 1)}{2} \left[ \left( \fdb \right)^2 + \fkbb \right]
  \right\} \,. 
\end{eqnarray}
\end{subequations}
Note, however, that the actual three-dimensional energy density does
\textit{not} evolve in the standard manner:
\begin{equation}
  \rho^{(3)} = V_{(n)} \rho \propto a^{-3(1 + w_a)} b^{-n w_b} \,, 
\end{equation}
and there is no reason to expect that $\rho_{\textrm{eff}}^{(3)}$
would either, if at the same time we want to keep the standard behavior
of $a$. Finally, an era of effective radiation (matter) domination
corresponding to $a \sim t^{1/2}$ ($a \sim t^{2/3}$)
and $w_{\textrm{eff}} \sim 1/3$ ($w_{\textrm{eff}} \sim 0$) need not
correspond to actual radiation (matter) domination, i.e.~$w_a \gg w_b$
($w_b \gg w_a$).

Finally, we would like to mention that according to Eq.~(\ref{mpl})
one expects the gravitational coupling constant $M_{Pl}^2$ to vary
with a time-dependent $V_{(n)} \sim b^n$. Since in the UED model all
particles are allowed to propagate in all dimensions, a similar case
can be made for other interactions. Therefore, any nonstatic solution
for $b$ must obey the tight observational bounds on the allowed
cosmological variation of the gravitational and electromagnetic
coupling constants (see, for example, \cite{uzan,mart,cli} and
references therein).

\section{Solutions with static extra dimensions}
\label{static}
Static extra dimensions is the only case considered so far in the UED
context and we saw above that there may be severe problems reproducing
standard cosmology otherwise. Let us therefore study whether the
field equations (\ref{allfe}) admit static solutions for $b$. Taking
the difference of twice Eq.~(\ref{fepq}) and the sum
of Eqs.~(\ref{fe00}) and three times~(\ref{feij}) gives
\begin{equation}
  \fddb + 3\fda \fdb + (n - 1) \left( \fdb \right)^2 + (n - 1) \fkbb +
  \frac{3w_a - 2w_b - 1}{n+2} \kappa^2 \rho = 0 \,. \label{bdot}
\end{equation}
From this we can immediately read off a necessary condition
for exactly static extra dimensions:
\begin{equation}
  (n + 2) (n - 1) \frac{k_b}{b^2} = (1 - 3w_a + 2w_b) \kappa^2
  \rho \,. \label{staticcondition} 
\end{equation}

If the extra dimensions are flat (for $n = 1$, the curvature is
automatically zero), this requires the universe to be empty ($\rho =
0$) or the equations of state to satisfy the following constraint:
\begin{equation}
  1 - 3w_a + 2w_b = 0 \,, \label{flatstat}
\end{equation}
which agrees with the five-dimensional models considered in
\cite{ell,kan}. In both cases, setting $\ddb = \db = 0$ reduces
Eqs.~(\ref{fe00}) and~(\ref{feij}) to the ordinary Friedmann equations
for a \dims-\,dimensional universe and Eq.~(\ref{fepq}) to a linear
combination of these. The static solutions are therefore consistent
with the full set of field equations. The particular combination $w_a
= 1/3$, $w_b = 0$ has been found and verified before
\cite{daemi,sai,gu}. For flat extra dimensions, there are thus two
ways of getting static solutions -- although the case of an empty
universe is, of course, not particularly interesting.

If, on the other hand, the extra dimensions are curved,
Eq.~(\ref{staticcondition}) requires $\rho$ to be constant for static
$b$.\footnote{Strictly speaking, this is true only for constant $w_a$,
which implies a constant $w_b$ according to Eq.~(\ref{eos}). From
standard cosmology, however, we expect long periods of approximately
constant $w_a$ (for example, during matter domination).} Unless $a$ is
also static, Eq.~(\ref{econserv}) then implies $w_a = -1$. The origin
of such an energy density could, for example, be a \ndims-\,dimensional
cosmological constant $\Lambda$, for which $\rho = \Lambda / \kappa^2$
and $w_a = w_b = -1$. Setting $\ddb = \db = 0$ then reduces
Eqs.~(\ref{allfe}) to the ordinary Friedmann equations for a
\dims-\,dimensional de Sitter universe, with an effective energy
density
\begin{equation}
  \rho^{(3)}_{\mathit{\textrm{eff}}} = \frac{M^2_{Pl}}{8\pi} \left(
  \Lambda -\frac{n(n - 1)}{2} \frac{k_b}{b^2} \right) =
  \frac{M^2_{Pl} \Lambda}{4\pi(n + 2)} \,,
\end{equation}
where the second equality follows from
Eq.~(\ref{staticcondition}). Curved, static extra dimensions are thus
possible in principle, but only for constant $\rho$. 
 
\section{Radiation domination}
\label{rad_dom}
Recent measurements of the cosmic microwave background indicate that
the universe is flat to a high degree of accuracy \cite{map}. We
therefore set the 3D curvature to zero. Moreover, we have shown that
static extra dimensions are incompatible with extra-dimensional
curvature in the case of non-negative pressure in 3D. Although the
latest results from type Ia supernovae observations strongly suggest
that the presently dominating energy component indeed \textit{does}
have negative pressure \cite{rie}, such a component, be it a
cosmological constant or a quintessence field, is believed to be
negligible until relatively recently and it therefore cannot provide
static, curved extra dimensions at earlier times. Thus we take $k_a =
k_b = 0$ from here on.

Shortly after freeze-out the LKP contribution should be negligible and the total energy density of the universe is therefore dominated by
ordinary radiation, i.e.,~the KK tower is largely unpopulated compared to relativistic particles with no
extra-dimensional momentum. Equation (\ref{eos}) then implies that the
extra-dimensional pressure should be negligible. Now, combining
Eqs. (\ref{eos}) and (\ref{flatstat}) we find that $w_a = 1/3$, $w_b =
0$ is the \textit{only} choice for the equations of state in a
universe that is dominated by relativistic particles and has exactly
static UED:
\begin{equation}
  b(t) = b_i \,. \label{b0}
\end{equation}
Furthermore, we have already seen that this choice reproduces standard
cosmological radiation domination in that the scale factor and energy 
density evolve as
\begin{eqnarray}
  a(t) &=& a_i \left( \frac{t}{t_i} \right)^\frac{1}{2} \,, \label{a0} \\
  \rho(t) &=& \rho_i \left( \frac{a(t)}{a_i} \right)^{-4} \,. \label{rho0}
\end{eqnarray}
Thus in the UED scenario, static extra dimensions arise naturally
during radiation domination.

However, from our discussion on pressure, we do not expect $w_a$ and
$w_b$ to take exactly these values and we must therefore consider
perturbations of the static solution. So let us assume that $0 \neq
w_b \ll 1$ and look for solutions of the form
\begin{subequations}
  \label{pert}
\begin{eqnarray}
  a(t) &=& a_*(t) + \deltaa(t) \,, \\
  b(t) &=& b_* + \deltab(t) \,, \\
  \rho(t) &=& \rho_*(t) + \deltarho(t) \,, 
\end{eqnarray}
\end{subequations}
where a star denotes the unperturbed solutions (\ref{b0})--(\ref{rho0})
and $|\deltaa|/a_*, |\deltab|/b_*, |\deltarho|/\rho_* \ll 1$, with
corresponding relations for the time derivatives of these
quantities. The linearized versions of the field
equations~(\ref{allfe}) are then given by
\begin{subequations}
\begin{eqnarray}
  \left( \frac{\da_*}{a_*} \right)^2 \left( 1 + 2\frac{\ddeltaa}{\da_*}
  - 2\frac{\deltaa}{a_*} \right) + n\frac{\da_*}{a_*}
  \frac{\ddeltab}{b_*} &=& \frac{\kappa^2}{3} \rho \,, \\
  2\frac{\dda_*}{a_*} \left( 1 +\frac{\dddeltaa}{\dda_*} -
  \frac{\deltaa}{a_*} \right) + \left( \frac{\da_*}{a_*} \right)^2
  \left( 1 + 2\frac{\ddeltaa}{\da_*} - 2\frac{\deltaa}{a_*} \right)
  + n \frac{\dddeltab}{b_*} + 2n \frac{\da_*}{a_*}\frac{\ddeltab}{b_*}
  &=& -\frac{\kappa^2}{3} (1 - nw_b) \rho \,, \\
  \frac{\dda_*}{a_*} \left( 1 + \frac{\dddeltaa}{\dda_*}
  - \frac{\deltaa}{a_*} \right) + \left( \frac{\da_*}{a_*} \right)^2
  \left( 1 + 2\frac{\ddeltaa}{\da_*} - 2\frac{\deltaa}{a_*} \right)
  + \frac{n - 1}{3} \frac{\dddeltab}{b_*} 
  + (n - 1)\frac{\da_*}{a_*}
  \frac{\ddeltab}{b_*} &=& -\frac{\kappa^2}{3}w_b \rho \,.
\end{eqnarray}
\end{subequations}
Subtracting the unperturbed equations this can be rewritten as
\begin{subequations}
\begin{eqnarray}
  2\left( \frac{\da_*}{a_*} \right)^2 \left( \frac{\ddeltaa}{\da_*} -
  \frac{\deltaa}{a_*} \right) + n\frac{\da_*}{a_*}
  \frac{\ddeltab}{b_*} &=& \frac{\kappa^2}{3}\deltarho \,,
  \label{uncoupled1} \\ 
  \frac{\dda_*}{a_*} \left( \frac{\dddeltaa}{\dda_*} -
  \frac{\deltaa}{a_*} \right) + 2\left( \frac{\da_*}{a_*} \right)^2
  \left( \frac{\ddeltaa}{\da_*} - \frac{\deltaa}{a_*} \right) &=&
  -\frac{\kappa^2}{3} nw_b \rho_* \,, \label{uncoupled2} \\
  \frac{\dddeltab}{b_*} + 3\frac{\da_*}{a_*} \frac{\ddeltab}{b_*} &=&
  \kappa^2 w_b \rho_* \,. \label{uncoupled3} 
\end{eqnarray}
\end{subequations} 
Equations~(\ref{uncoupled2}) and~(\ref{uncoupled3}) are uncoupled
differential equations for $\deltaa$ and $\deltab$ which can be
solved after inserting the unperturbed solutions (\ref{a0}),
(\ref{rho0}). Equation (\ref{uncoupled1}) can be interpreted as defining
$\deltarho$ and is thus automatically satisfied. The general solutions
are found to be:
\begin{subequations}
\label{general}
\begin{eqnarray}
  \deltaa(t) &=& A_1 \left( \frac{t}{t_i} \right)^{1/2} + A_2
  \left( \frac{t}{t_i} \right)^{-1/2} - 
  \frac{1}{3}  n \kappa^2\rho_i t_i^2 a_i\,
  \left(\frac{t}{t_i}\right)^{1/2} \int_{t_i}^t w_b(x)
  (x^{-1} - t^{-1}) \, dx \,, \\  
  \deltab(t) &=& B_1 + B_2 \left( \frac{t}{t_i} \right)^{-1/2}
  + 2 \kappa^2\rho_i t_i^2 b_i\,
  \int_{t_i}^t w_b(x)(x^{-1} - t^{-1/2}x^{-1/2}) \,dx \,,  
\end{eqnarray}
\end{subequations}
where $A_1$,$A_2$,$B_1$,$B_2$ are integration constants fixed by
the initial conditions at time $t_i$. Note that $A_1$ and $B_1$ can
just as well be regarded as part of the initial conditions for the
unperturbed solutions $a_*$ and $b_*$ respectively. Thus for $w_b = 0$
we find only decaying solutions for the perturbations in both scale
factors -- i.e.~the static solution given by Eqs.~(\ref{b0}),(\ref{a0})
is stable under small perturbations, as claimed previously (without
proof) in \cite{gu}.

In the case of $w_b \neq 0$, however, there might also exist a growing
solution.  For example, a constant $w_b$ gives\footnote{A more
realistic time dependence of $w_b$ is considered in
Sec.~\ref{trans}, where we find a more rapid growth of $\deltab$.}
\begin{subequations}
\begin{eqnarray}
  \deltaa(t) \sim - \frac{1}{3} w_b n \kappa^2\rho_i t_i^2 a_i\,
  \left(\frac{t}{t_i}\right)^{1/2}
  \ln\left(\frac{t}{t_i}\right), \\   
  \deltab(t) \sim 2 w_b \kappa^2\rho_i t_i^2 b_i\,
  \ln\left(\frac{t}{t_i}\right).
\end{eqnarray}
\end{subequations}
Since the relative perturbations of the
scale factors grow only logarithmically in both cases,
\begin{equation}
  \frac{|\deltaa|}{a_*}, \frac{|\deltab|}{b_*} \propto \ln t \,,
\end{equation}
these perturbations can still be expected to remain relatively small
during radiation domination after LKP freeze-out.

\section{Matter domination}
\label{mat_dom}
We have shown that there are approximately static solutions for the
UED during radiation domination. In fact, we found that $w_a = 1/3$
and $w_b = 0$ is the only possible choice for the equations of state
that can give exactly static, flat extra dimensions in a universe
dominated by relativistic particles. However, as we remarked before,
the universe is \textit{always} dominated by relativistic particles if
a significant amount of the dark matter is made up of LKPs. It is
therefore clear that there are \textit{no} exactly static solutions
for the UED during matter domination, i.e.~$w_a \approx 0$.

Having excluded exactly static extra dimensions, the next case of interest
would be slowly evolving solutions:
\begin{equation}
  \left| \fdb \right| \ll \left| \fda \right| \,. \label{slow}
\end{equation}
But what ansatz should we make for $a$? A significant, long-term
deviation from the usual time evolution during matter domination would
most likely alter the predictions of standard cosmology concerning,
e.g., large-scale structure formation. Let us therefore first examine
whether there are any solutions to the field equations that give
\begin{equation}
  a(t) = a_\star(t) \equiv a_i \left( \frac{t}{t_i}
  \right)^\frac{2}{3} \,. \label{md_a} 
\end{equation}
We noted before that $w_a = 0$ implies $w_b = 1/n$. The
field equations~(\ref{allfe}) can then be rewritten as two homogeneous
equations for $\dda$ and $\ddb$ and one defining equation for
$\rho$. With the above expression for $a$ we get
\begin{subequations}
\begin{eqnarray}
  \left( \frac{\da_\star}{a_\star} \right)^2 + n
  \frac{\da_\star}{a_\star} \fdb + \frac{n(n - 1)}{6}
  \left( \fdb \right)^2 &=& \frac{\kappa^2}{3} \rho, \label{md_fe1} \\
  \fddb + 2 \frac{\da_\star}{a_\star} \fdb + \frac{n -
  1}{2} \left( \fdb\right)^2 &=& 0 \,, \label{md_fe2} \\
  \frac{3}{2n} \frac{\da_\star}{a_\star} + \fdb &=&
  0 \,. \label{md_fe3} 
\end{eqnarray}
\end{subequations}
Differentiating Eq.~(\ref{md_fe3}), one finds that the last two
equations are inconsistent with each other for $n \neq -3$, so there
are \textit{no} solutions with $a(t) = a_\star(t)$ and $w_a
= 0$. Neither can there be any solutions with $w_a \approx 0$
and $a$ of the form
\begin{equation}
  a(t) = a_\star(t) + \deltaa(t) \,,
\end{equation}
where $|\deltaa|/a_\star, |\ddeltaa|/\da_\star,
|\dddeltaa/\dda_\star| \ll 1$. This is because, to zeroth order,
inserting such an ansatz would leave the unperturbed equations
unchanged.  Of course, if we allow for rapidly oscillating $a(t)$ then
$\ddeltaa$ and $\dddeltaa$ need not be small. We do not consider such
behavior here.

So are there any solutions to the field equations at all during matter
domination? With the quite general ansatz
\begin{equation}
  a(t) = a_i \left( \frac{t}{t_i} \right)^x \,,
\end{equation}
we find that all solutions are of the form
\begin{equation}
  b(t) = b_i \left( \frac{t + B}{t_i + B} \right)^y \,.
\end{equation} 
Some of these, namely
\begin{equation}
\label{xy}
\begin{cases}
  x &= \frac{3 \pm \sqrt{3n(n + 2)}}{3(n + 3)}, \\
  y &= \frac{n \mp \sqrt{3n(n + 2)}}{n(n + 3)}, \\
  B &\equiv 0,
\end{cases}
\end{equation}
are vacuum solutions. In fact, they are known as Kasner-type solutions
and have been found before under the \textit{assumption} of an empty
universe and a power-law behaviour of both scale factors
\cite{cho}. The only additional solutions appear when $n \neq 1$ and
are given by
\begin{equation}
\label{xy2}  
\begin{cases} 
  x & = 0, \\
  y & = \frac{2}{n + 1},
\end{cases}
\end{equation}
with $B$ being an arbitrary integration constant. Although they describe a
nonempty universe they have a static scale factor $a$. Of course, for
$n = 3$ we knew of this solution beforehand, since $w_a = 0$ and $w_b
= \frac{1}{3}$ is the static, radiation dominated solution with $a$
and $b$ interchanged.

With no suitable solutions describing actual matter domination ($w_a
\ll w_b$), we turn to the possibility of an era of effective matter
domination, i.e.,~$p^{(3)}_{\textrm{eff}} = 0$. Inserting the defining
Eqs.~(\ref{effective}) into the field equations~(\ref{allfe}) then
yields the familiar result $a \propto t^{2/3}$ by
construction. Using Eq.~(\ref{eos}) we can write the remaining equations
as
\begin{subequations}
\begin{eqnarray}
  \fddb &=& -\frac{2}{3nt^2} - \frac{2}{t} \left( \fdb \right) -
  \frac{n - 1}{2} \left( \fdb \right)^2 \,, \label{diffb} \\
  w_a &=& \frac{4(1 + nt\fdb)}{8 + 12nt\fdb + 3n(n -1)t^2 \left( \fdb
  \right)^2} \,, \\
  \kappa^2 \rho &=& \frac{4}{3t^2} + \frac{2n}{t} \fdb + \frac{n(n -
  1)}{2} \left( \fdb \right)^2 \,.
\end{eqnarray}
\end{subequations}
Equation (\ref{diffb}) has the general solution
\begin{equation}
  b = B_1 \left[ \left( \frac{t}{t_i} \right)^{-1} \cos^{\,2} \left(
  \sqrt{\frac{n + 4}{12n}} \ln \frac{t}{t_i} + B_2 \right)
  \right]^{1/(n + 1)} \,, \label{bounce} 
\end{equation}
where $B_1$ and $B_2$ are integration constants to be fixed by the
initial conditions. This corresponds to decaying, bouncing extra
dimensions. Although such a behavior of $b$ might be possible, the
corresponding evolution of $w_a$ is not -- in fact it is
singular. Indeed, in our model there is no physical motivation for why
$p^{(3)}_{\textrm{eff}}$ should vanish for nonstatic $b$, and it is therefore
no surprise that we get unphysical solutions from imposing it.

\section{Transition Period}
\label{trans}
So far we have focused on the two extreme cases of having zero
pressure in either the UED (radiation domination) or 3D (matter
domination). In a final attempt to find solutions which resemble the
standard cosmological evolution of $a$ (in particular during matter
domination), we will now make a more general numerical study of the
transition from an era of radiation domination with approximately
static UED to one with a sizable energy density contribution from
LKPs. In order to do this we make the approximation that the LKPs
have only extra-dimensional momentum, which should be valid for
temperatures below $\sim 1~\textrm{TeV}$. This allows us to split the
energy density and pressure into two parts:
\begin{eqnarray}
  \rho &=& \rho_r + \rho_m \,, \\
  p_a &=& w_a^r \rho_r + w_a^m \rho_m = \frac{\rho_r}{3} \,, \\
  p_b &=& w_b^r \rho_r + w_b^m \rho_m = \frac{\rho_m}{n} \,, \label{p_b}
\end{eqnarray}
where $r$ and $m$ denote ordinary particles (radiation) and LKPs
(matter), respectively. Neglecting interactions, the energy-momentum is
separately conserved and Eq.~(\ref{econserv}) gives:
\begin{subequations}
\label{rm_con}
\begin{eqnarray}
  \rho_r &=& \rho_{r_i} \left( \frac{a}{a_i} \right)^{-4} \left(
  \frac{b}{b_i} \right)^{-n} \,, \\
  \rho_m &=& \rho_{m_i} \left( \frac{a}{a_i} \right)^{-3} \left(
  \frac{b}{b_i} \right)^{-(n + 1)} \,.
\end{eqnarray} 
\end{subequations}
Now introduce the dimensionless variable $t^\prime \equiv
t/t_i$ and rescale
\begin{equation}
  a \rightarrow \frac{a}{a_i}, \qquad b \rightarrow \frac{b}{b_i} \,.
\end{equation}
Using a prime to denote differentiation with respect to $t^\prime$,
Eqs.~(\ref{feij}),(\ref{fepq}) become
\begin{subequations}
\label{allfe'}
\begin{eqnarray}
  2\fppa + \left( \fpa \right)^2 + n \fppb + 2n \fpa \fpb + \frac{n(n
  - 1)}{2} \left( \fpb \right)^2 &=& -\frac{\kappa^2 t_i^2
  \rho_{r_i}}{3 a^4 b^n} \,, \label{fe2} \\
  3\fppa + 3\left(\fpa \right)^2 + (n-1)\fppb + 3(n-1)\fpa \fpb +
  \frac{(n - 1)(n - 2)}{2} \left( \fpb \right)^2 &=& -\epsilon 
  \frac{\kappa^2 t_i^2 \rho_{r_i}}{n a^3 b^{n + 1}} \,, \label{fe3}
\end{eqnarray}
\end{subequations}
where $\epsilon \equiv \rho_{m_i}/\rho_{r_i}$, and from
Eq.~(\ref{fe00}) we get
\begin{equation}
  \kappa^2 t_i^2 \rho_{r_i} = \frac{3}{1 +\epsilon} \left[ \left( \fpa
  \right)^2 + n \fpa \fpb + \frac{n(n - 1)}{6} \left( \fpb \right)^2
  \right]_{t^\prime = 1} \,.
\end{equation}

Starting from the solutions (\ref{pert}),(\ref{general}) for radiation
domination and approximately static extra dimensions, the appropriate
initial conditions are given by
\begin{eqnarray}
\label{initial}
  a(1) = 1 \,, &\quad& \pa(1) = \frac{1}{2} - 
  \frac{n}{3} \epsilon\,, \nonumber \\
  \quad b(1) = 1 \,, &\quad& \pb(1) = \frac{3}{4} \epsilon \,, \nonumber \\
  &\epsilon \ll 1\,.&  
\end{eqnarray}
Here, we keep track of terms linear in $\epsilon$ in order to be
consistent with the expected behavior of slowly \textit{growing}
extra dimensions (as opposed to the case of exactly static extra
dimensions that would result from $\epsilon\equiv 0$). Of course, one
could in principle imagine different initial conditions, but the ones
chosen above correspond naturally to the setup presented here and in
Sec.~\ref{rad_dom}.

\begin{figure}[t!]
  \psfrag{a}[][][1.4]{$a$}
  \psfrag{b}[][][1.4]{$b$}
  \psfrag{r}[][][1.4]{$\frac{\rho_m}{\rho}$}
  \psfrag{x}[][][1.4]{$\log t'$}
  \psfrag{y1}[][][1.4][90]{$\log a,~\log b$}
  \psfrag{y2}[][l][1.4][90]{$\rho_m/\rho$}
  \psfrag{0}[][][1.1]{$0$}
  \psfrag{2}[][][1.1]{$2$}
  \psfrag{4}[][][1.1]{$4$}
  \psfrag{6}[][][1.1]{$6$}
  \psfrag{8}[][][1.1]{$8$}
  \psfrag{5}[][][1.1]{$5$}
  \psfrag{10}[][][1.1]{$10$}
  \psfrag{15}[][][1.1]{$15$}
  \psfrag{20}[][][1.1]{$20$}
  \psfrag{0.0}[r][][1.1]{$0.0$}
  \psfrag{0.1}[r][][1.1]{$0.1$}
  \psfrag{0.2}[r][][1.1]{$0.2$}
  \psfrag{0.3}[r][][1.1]{$0.3$}
  \psfrag{0.4}[r][][1.1]{$0.4$}
  \psfrag{0.5}[r][][1.1]{$0.5$}
  \psfrag{0.6}[r][][1.1]{$0.6$}
  \psfrag{0.7}[r][][1.1]{$0.7$}
  \includegraphics[width=.9\columnwidth]{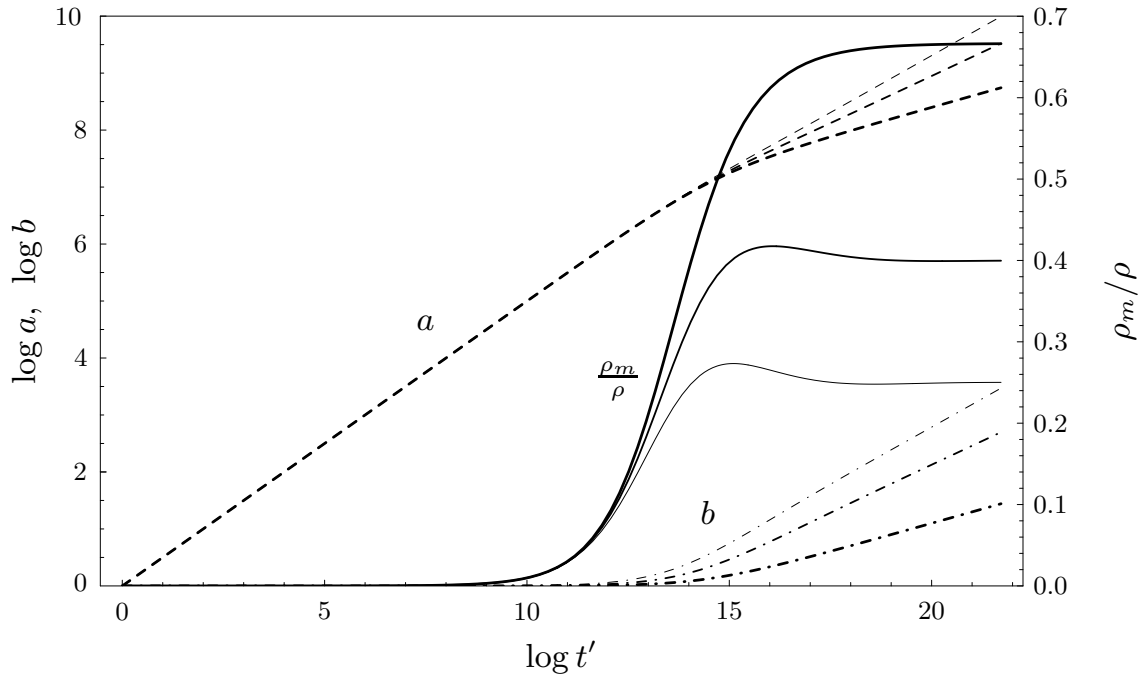}
  \caption{\label{eps_5} \footnotesize{The evolution of the scale
  factors $a$ (dashed) and $b$ (dash-dotted), as well as the LKP
  energy density $\rho_m/\rho$ (solid) for $n =$1 (thin), 2
  (medium), and 7 (thick) with $\epsilon\equiv 
  \rho_{m_i} / \rho_{r_i} = 10^{-7}$. For
  $\rho_m/\rho\lesssim 0.1$ the extra dimensions are nearly
  static and the evolution of $a$ reproduces the radiation dominated
  regime of standard cosmology to a very good approximation. For
  $\rho_m/\rho> 0.1$ however, neither $a$ nor $b$ show
  the desired behaviour.}}
\end{figure}
\begin{figure}[ht!]
  \psfrag{a}[][][1.4]{$a$}
  \psfrag{b}[][][1.4]{$b$}
  \psfrag{r}[][][1.4]{$\frac{\rho_m}{\rho}$}
  \psfrag{x}[][][1.4]{$\log t'$}
  \psfrag{y1}[][][1.4][90]{$\log a,~\log b$}
  \psfrag{y2}[][l][1.4][90]{$\rho_m/\rho$}
  \psfrag{0}[][][1.1]{$0$}
  \psfrag{2}[][][1.1]{$2$}
  \psfrag{4}[][][1.1]{$4$}
  \psfrag{6}[][][1.1]{$6$}
  \psfrag{8}[][][1.1]{$8$}
  \psfrag{5}[][][1.1]{$5$}
  \psfrag{10}[][][1.1]{$10$}
  \psfrag{15}[][][1.1]{$15$}
  \psfrag{20}[][][1.1]{$20$}
  \psfrag{0.0}[r][][1.1]{$0.0$}
  \psfrag{0.1}[r][][1.1]{$0.1$}
  \psfrag{0.2}[r][][1.1]{$0.2$}
  \psfrag{0.3}[r][][1.1]{$0.3$}
  \psfrag{0.4}[r][][1.1]{$0.4$}
  \psfrag{0.5}[r][][1.1]{$0.5$}
  \psfrag{0.6}[r][][1.1]{$0.6$}
  \psfrag{0.7}[r][][1.1]{$0.7$}
  \includegraphics[width=.9\columnwidth]{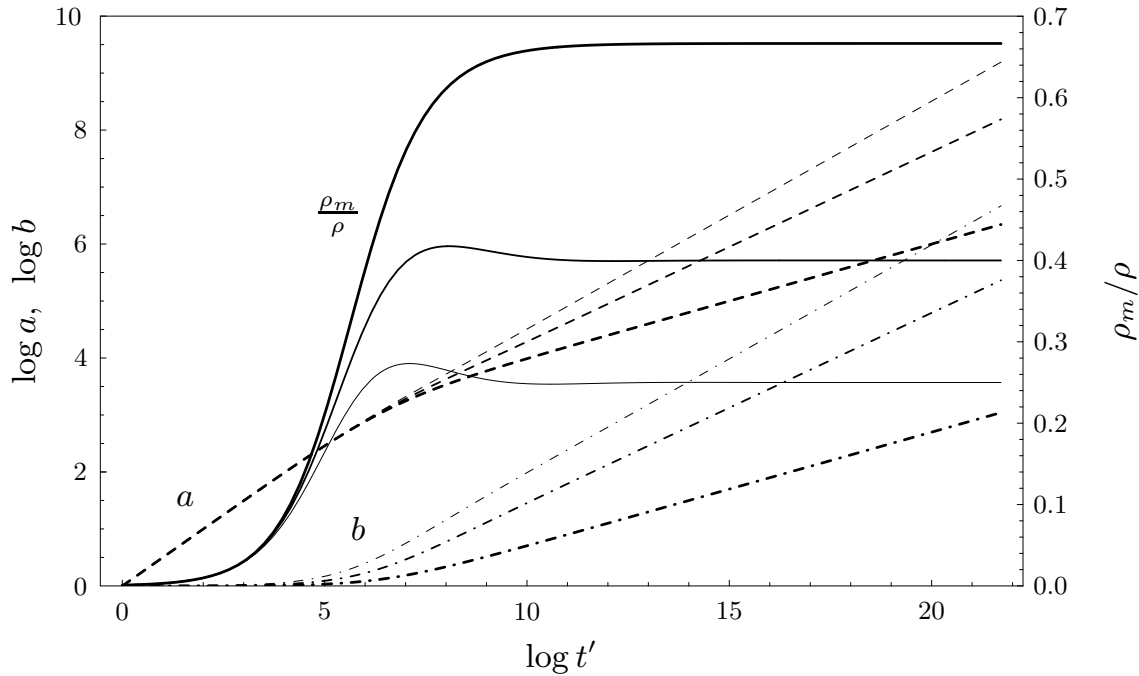}
  \caption{\label{eps_3} \footnotesize{This plot shows the evolution of 
  the same quantities as in Fig. 1, this time with $\epsilon =
  10^{-3}$. As expected, a larger value of $\epsilon$ gives a shorter period of
  radiation domination.}}
\end{figure}

The numerical solutions to Eqs.~(\ref{allfe'}) are plotted in
Figs.~\ref{eps_5} and~\ref{eps_3} for different numbers of extra
dimensions and values of $\epsilon$. In the beginning, we find the
behavior expected from our discussion in Sec.~\ref{rad_dom}
\footnote{With the explicit time dependence of $w_b$ given by
Eq.~(\ref{p_b}), we find that $\deltab$ grows as $t^{1/2}$,
which is much faster than logarithmically. The radiation dominated era
is therefore shorter than expected from Eqs.~(\ref{general}), although
the qualitative behavior remains the same.} -- very slowly growing
extra dimensions and an expansion of 3D that corresponds to the usual
radiation domination $a \propto t^{\prime 1/2}$. However, as
soon as the LKPs make up roughly 10\% of the total energy density of
the universe, the extra dimensions start to expand at a rate
comparable to $a$. Such a rapid expansion of the UED is already ruled
out by present bounds on the time variation of the electromagnetic and
gravitational coupling constants (see,
e.g.,~\cite{uzan,mart,cli}). Moreover, when the extra dimensions are
increasing, our scale factor increases \textit{less} rapidly than
$t^{\prime 1/2}$ -- instead of approaching $t^{\prime 2/3}$ as
predicted by standard cosmology.

The behavior of the scale factors as described above is stable
against perturbations in the initial conditions of $\pa$ and $\pb$ of
the order of $\epsilon$. Allowing for even larger perturbations, the
only qualitatively different behavior we find is collapsing $b$ and
collapsing or inflating $a$ - these solutions are obviously not viable
alternatives either.

\section{Discussion and Conclusions}
\label{conc}
The identity of dark matter is one of the most challenging puzzles in
modern particle physics and cosmology. Recently, it has been noted
that models with universal extra dimensions provide a natural WIMP
candidate that could make up a significant amount of the dark matter
today. This subsequently led to a great deal of interest in studying the 
detection properties of these particles. However, the estimates for
today's LKP abundance depend crucially on the underlying assumption of 
static extra dimensions and a standard cosmological evolution history of 
the universe since the time of freeze-out of these particles.

In this article we have studied in detail whether one can expect such
 behavior without adding an explicit stabilization mechanism. To
this end we have analyzed cosmological solutions to Einstein's field
equations in \ndims dimensions that are appropriate to describe a
universe with UED. More specifically, our setup is given by a
homogeneous FRW metric with two different scale factors and the
assumption that LPKs make up the dominant part of the dark matter,
i.e.,~the universe is always dominated by relativistic particles.

We find that a natural -- and in fact the only -- way to get exactly
static extra dimensions in this scenario is to set $w_a = \frac{1}{3}$
and $w_b = 0$. This also reproduces the usual radiation-dominated
behavior of the scale factor in 3D. Allowing for $0 \neq w_b \ll 1$,
which is much more realistic in the UED scenario, we still find
approximately constant extra dimensions and $a \sim
t^{1/2}$.

However, during matter domination ($w_a = 0$) there are no static
solutions for the extra dimensions. Even worse, there are no solutions
at all that are consistent with the standard matter dominated
behavior of the scale factor in 3D, $a(t) \propto
t^{2/3}$. With a more general ansatz $a(t) \propto t^x$ we do
find solutions, but for $x \neq 0$ they all describe an empty
universe. Demanding $w_{\textrm{eff}}=0$ instead of $w_a=0$ we get the
usual behavior of $a(t)$ by construction, but the corresponding
solutions for $b(t)$ and $\rho(t)$ are unphysical.

In the reasonable approximation that the LKPs have only
extra-dimensional momentum, we have also performed a numerical
analysis of the transition from a radiation dominated universe with
approximately static UED to one with a sizable energy density
contribution from LKPs. The evolution of $b$ is generically found to
be much too rapid given the present bounds on the time variation of
the electromagnetic and gravitational coupling constants, and $a$ does
not show the standard behavior either.

To summarize, we have shown that, within our framework, an explicit
mechanism is needed not only to stabilize the UED but also to
reproduce standard cosmology in 3D during matter domination.  Although
one could consider more complex models, e.g.,~with different scale
factors for each extra dimension, we believe that finding static
solutions -- or indeed any solutions which give
standard cosmology for both radiation and matter domination without
obviously violating experimental bounds on the evolution of the extra
dimensions -- is a generic difficulty of this scenario.

\begin{acknowledgments}
We are grateful to Lars Bergstr\"om, Joakim Edsj\"o, Anne Green,
Stefan Hofmann and Edvard M\"ortsell for helpful discussions and
careful reading of the manuscript. 
\end{acknowledgments} 


\end{document}